\documentclass[journal,twoside,web]{FrontBack/ieeecolor}
\usepackage{FrontBack/tuffc}
\usepackage{algorithmic}
\usepackage{textcomp}
\usepackage{wrapfig,colortbl}
\definecolor{abstractbg}{rgb}{1,0.969,0.914}
\usepackage{markdown}
\usepackage[utf8]{inputenc}
\usepackage{array}
\usepackage{amsmath, amssymb,amsfonts}
\usepackage{expl3,graphicx}
\usepackage{mwe}
\usepackage{adjustbox}
\usepackage{xparse}
\usepackage[normalem]{ulem} 
\usepackage{svg}

\usepackage{tikz}

\newcommand{\rSrc}{\mathbf{S}}
\newcommand{\rRcv}{\mathbf{R}}
\newcommand{\vSrc}{\rSrc'}
\newcommand{\vRcv}{\rRcv'}
\newcommand{\resLim}[1]{\varepsilon_\text{#1}}

\newcommand{\timeshift}{\tau}

\newcommand{\roi}{\text{ROI}}

\NewDocumentCommand{\tMat}{s m O{}}{{
    \IfValueTF{#1}
    {\mathbf{\tau}}
    {\mathbf{\tau}}
}^{#2}_{#3}}
\newcommand{\model}{\text{model}}
\newcommand{\meas}{\text{meas}}
\NewDocumentCommand{\residual}{O{}}{e_{#1}}
\newcommand{\misfit}{m}

\newcommand{\sos}[1]{\mathbf{c}^#1}
\newcommand{\cin}{c_\text{ROI}}
\newcommand{\cout}{c_\text{ER}}

\newcommand{\errorfreeleft}{\left(}
\newcommand{\errorfreeright}{\right)}

\newcommand{\forwardProp}{\rightarrow}
\newcommand{\backProp}{\rightsquigarrow}
\NewDocumentCommand{\sig}{m o o o s o o}{ 
    {
    \text{sig}_{
        \IfNoValueTF{#1}
            { }
            {#1}
        \IfNoValueTF{#2}
            {}
            {\forwardProp #2}
        \IfNoValueTF{#3}
            {}
            {\forwardProp #3}
        \IfNoValueTF{#4}
            {}
            {\forwardProp #4}
        \IfValueTF{#5}
            {
                \IfNoValueTF{#6}
                    {}
                    {\backProp #6}
                \IfNoValueTF{#7}
                    {}
                    {\backProp #7}
            }
            {}
        }        
    }
}

\NewDocumentCommand{\prop}{m o o o s o o}{ 
    {
    \mathfrak{A}_{
        \IfNoValueTF{#1}
            { }
            {#1}
        \IfNoValueTF{#2}
            {}
            {\forwardProp #2}
        \IfNoValueTF{#3}
            {}
            {\forwardProp #3}
        \IfNoValueTF{#4}
            {}
            {\forwardProp #4}
        \IfValueTF{#5}
            {
                \IfNoValueTF{#6}
                    {}
                    {\backProp #6}
                \IfNoValueTF{#7}
                    {}
                    {\backProp #7}
            }
            {}
        }        
    }
}

\newcommand{\relLocSym}{\mathbf{r}}
\NewDocumentCommand{\relLoc}{t^ m o o o s o o}{
    { 
        {\IfBooleanTF {#1} 
        {\hat{\relLocSym}}
        {\relLocSym}}
        _{
            \IfNoValueTF{#2}
                { }
                {#2}
            \IfNoValueTF{#3}
                {}
                {\forwardProp #3}
            \IfNoValueTF{#4}
                {}
                {\forwardProp #4}
            \IfNoValueTF{#5}
                {}
                {\forwardProp #5}
            \IfValueTF{#5}
                {
                    \IfNoValueTF{#6}
                        {}
                        {\backProp #6}
                    \IfNoValueTF{#7}
                        {}
                        {\backProp #7}
                }
                {}
        }
    }
}

\newcommand{\poiSym}{\zeta}

\NewDocumentCommand{\poi}{m o o o s o o}{ 
    {
        \poiSym_
        {\errorfreeleft
            \IfNoValueTF{#1}
                { }
                {#1}
            \IfNoValueTF{#2}
                {}
                {\forwardProp #2}
            \IfNoValueTF{#3}
                {}
                {\forwardProp #3}
            \IfNoValueTF{#4}
                {}
                {\forwardProp #4}
            \IfValueTF{#5}
                {
                    \IfNoValueTF{#6}
                        {}
                        {\backProp #6}
                    \IfNoValueTF{#7}
                        {}
                        {\backProp #7}
                }
                {}
        \errorfreeright}
    }
}

\newcommand{\figref}[1]{Fig.~\ref{#1}}
\usepackage[backend=biber,style=ieee]{biblatex}
\svgpath{{Fig/}}
\graphicspath{{Fig/}}
\usepackage{hyperref} 
\hypersetup{
    colorlinks=true,
    linkcolor=magenta,
    citecolor=blue,
    linktocpage=true
}
\usepackage[figure,figure*]{hypcap}

\renewcommand{\eqref}[1]{(\ref{#1})}
\newcommand{\pareneqref}[1]{(\ref{#1})}

\begin{document}
\title{Virtual Extended-Range Tomography (VERT): Contact-free realistic ultrasonic bone imaging}
\author{Aaron Chung-Jukko, Peter Huthwaite
\thanks{A. Chung-Jukko is with the Non-Destructive Evaluation Research Group, Department of Mechanical Engineering, Imperial College London, London, United Kingdom SW7 2AZ (email: aaron@chung-jukko.com) }
\thanks{P. Huthwaite is with the Non-Destructive Evaluation Research Group, Department of Mechanical Engineering, Imperial College London, London, United Kingdom SW7 2AZ (e-mail: p.huthwaite@imperial.ac.uk)}}

\IEEEtitleabstractindextext{%
\fcolorbox{abstractbg}{abstractbg}{%
\begin{minipage}{\textwidth}\rightskip2em\leftskip\rightskip\bigskip
\begin{wrapfigure}[19]{r}{3.3in}%
\vspace{-1pc}
\hspace{-3pc}
\includegraphics[width=3.3in]{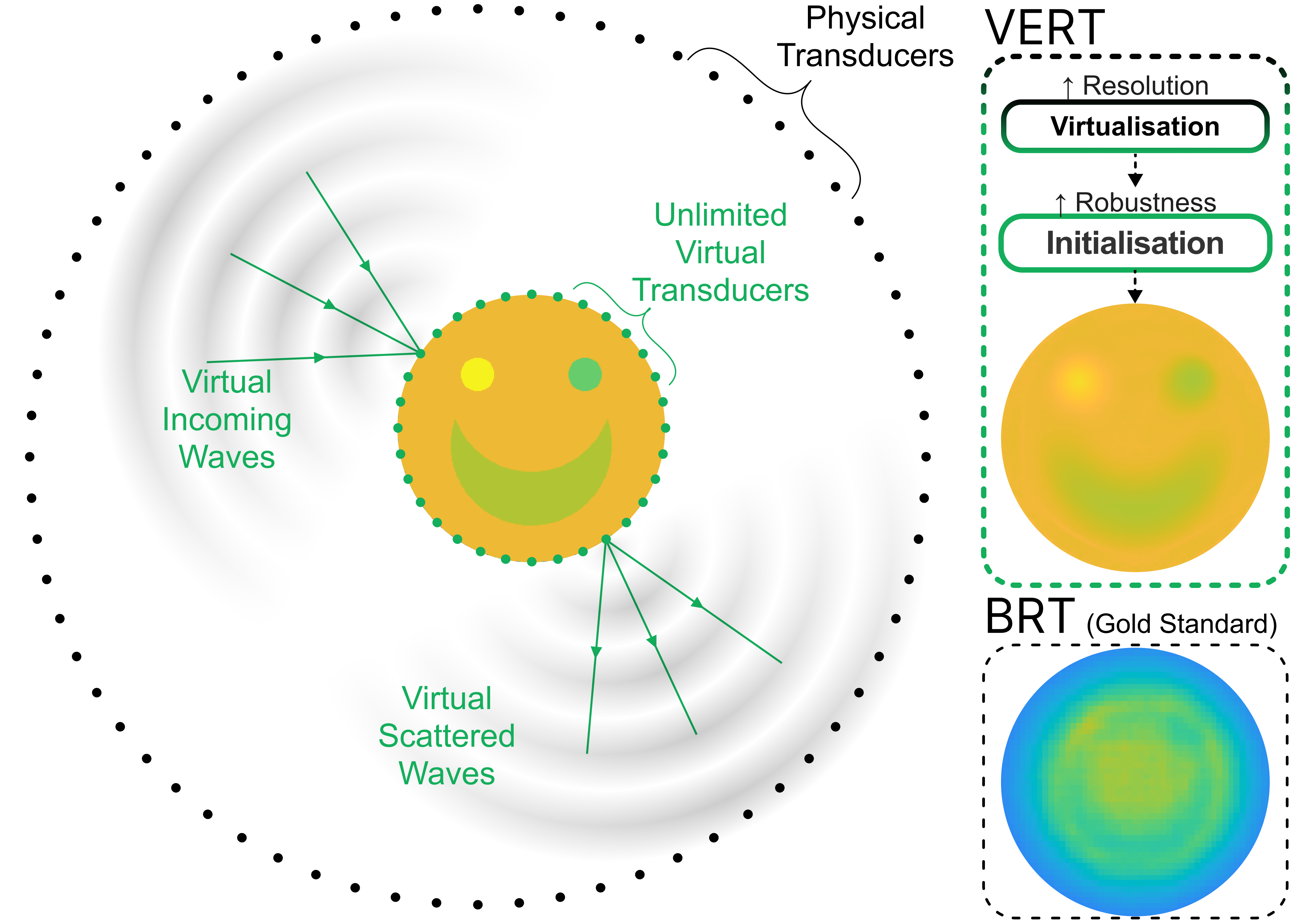}
\end{wrapfigure}%
\begin{abstract}
Ultrasound tomography generally struggles to reconstruct high-contrast and/or extended-range problems.
A prime example is site-specific \textit{in-vivo} bone imaging, crucial for accurately assessing the risk of life-threatening fractures, which are preventable given accurate diagnosis and treatment.
In this type of problem, two main obstacles arise:
(a) an external region prohibits access to the region of interest (ROI), and
(b) high contrast exists between the two regions.
These challenges impede existing algorithms---including bent-ray tomography (BRT), known for its robustness, speed, and reasonable short-range resolution. 
We propose Virtual Extended-Range Tomography (VERT), which tackles these challenges through:
(a) placement of virtual transducers directly on the ROI, facilitating
(b) rapid initialisation before BRT inversion.
\textit{In-silico} validation against BRT with and without \textit{a-priori} information shows superior resolution and robustness---while maintaining or even improving speed. 
These improvements are drastic where the external region is much larger than the ROI.
Additional validation against the practically impossible---BRT directly on the ROI---demonstrates that VERT is approaching the resolution limit.
The capability to solve high-contrast extended-range tomography problems without prior knowledge about the ROI's interior has many implications.
VERT has the potential to unlock site-specific \textit{in-vivo} bone imaging for assessing fracture risk, potentially saving millions of lives globally.
In other applications, VERT may replace classical BRT to yield improvements in resolution, robustness and speed---especially where the ROI does not cover the entire imaging array. 
For even higher resolution, VERT offers a reliable starting background to complement algorithms with less robustness and high computational costs.
\end{abstract}
\bigskip
\end{minipage}
}
}

\maketitle

\begin{table*}[!t]
\arrayrulecolor{subsectioncolor}
\setlength{\arrayrulewidth}{1pt}
{\sffamily\bfseries\begin{tabular}{lp{6.75in}}\hline
\rowcolor{abstractbg}\multicolumn{2}{l}{\color{subsectioncolor}{\itshape
Highlights}{\Huge\strut}}\\
\rowcolor{abstractbg}$\bullet$ & 
Virtual Extended-Range Tomography (VERT) addresses high-contrast extended-range problems by placing virtual transducers on the region of interest---from which bent-ray tomography is performed.\\
\rowcolor{abstractbg}$\bullet${\large\strut} & 
VERT drastically improves resolution, robustness, and speed versus classical bent-ray tomography.\\
\rowcolor{abstractbg}$\bullet${\large\strut} & 
Accurate \textit{in-vivo} site-specific bone imaging is within reach. In other applications, VERT may replace bent-ray tomography as the standard method.\\[2em]\hline
\end{tabular}}
\setlength{\arrayrulewidth}{0.4pt}
\arrayrulecolor{black}
\end{table*}

\section{Introduction} \label{sec:Introduction}
Ultrasonic tomography is widely applied in diverse fields from medicine to geophysics and engineering. Tomography reconstructs material properties across a region of interest (ROI), which are seldom directly accessible and must not be destroyed in the process. These reconstructions carry critical information, hence tomographic algorithms have to balance resolution, speed, and robustness. While resolution and speed are conceptually straightforward and typically quantifiable, robustness reflects the need for an algorithm which reliably produces an accurate image with limited sensitivity to undesired property variations, minimal manual tuning of the routine (e.g. through weighting factors), and good convergence properties.

One of the most robust and fastest algorithms is Bent-Ray Tomography (BRT), also known as Time-of-Flight Tomography or Travel Time Tomography \parencite{Huthwaite.2014}. Current applications of BRT range from corrosion mapping to breast imaging, and as a starting background for algorithms with higher resolution but lower robustness \parencite{Huthwaite.2014,LiFMM,HARBUT}. 
However, despite its robust behaviour, BRT struggles with increasing contrast and poor resolution. The latter arises from ignoring diffraction effects and is exacerbated when the ray paths are long. These drawbacks are amplified in high-contrast extended-range tomography problems, where the ROI has a high acoustic contrast with the external region (ER) and is also distant from the array.

An archetype of these problems is \textit{in-vivo} ultrasonic imaging of the proximal femur---the ROI (bone) is surrounded by soft tissues (the ER), resulting in high contrast and a large array diameter. This imaging is needed to determine the risk of hip fractures, the deadliest and costliest type of osteoporotic fracture \parencite{eu-osteo,causal}. The current diagnostic gold standard, dual X-ray absorptiometry, measures bone mineral density rather than strength and has high specificity but low sensitivity \parencite{who1994fracture}. Ultrasound-based methods are a promising alternative given their ability to uncover bone quality metrics such as porosity and stiffness \parencite{BoneQUS_Intro}, but existing tomographic methods ``are still too limited today to be used in clinical application" \parencite{BoneQUS_Chapter5}. These limitations have redirected significant research efforts towards the suboptimal approach of correlating fracture site properties from ultrasonic characterisation of vastly different peripheral skeletal sites \parencite{BoneQUS_Axial1,BoneQUS_Axial2}. There are also other examples outside medicine, such as in non-destructive testing with guided wave tomography, where it is desired to image a local inhomogeneity. An imaging algorithm that can successfully reconstruct high-contrast extended-range tomography is desperately needed.

In this paper, we propose a novel algorithm Virtual Extended-Range Tomography (VERT). The main goals are to enhance BRT's 1. extended-range resolution and 2. high-contrast robustness, validated in the context of bone imaging.

Achieving these goals holds the potential to significantly advance the accurate diagnosis, monitoring and treatment of osteoporosis, thereby impacting the lives of approximately one-third of women and one-fifth of men aged 50 and above \parencite{eu-osteo,long-term-risk}. VERT will also find applications outside bone imaging in most scenarios where BRT is applicable, with resolution gains achievable for most problems. The robustness and improved resolution of VERT will also provide an excellent starting background for other high-resolution but less robust and computationally intensive algorithms, such as full-waveform inversion.

The theory behind BRT's performance in high-contrast extended-range tomography is discussed in Section \ref{sec:BRTVERT}. Section \ref{sec:New Technique} explores the key innovations of VERT: virtualisation of transducers onto the ROI boundary and the initialisation step, which improves high-contrast robustness. In Section \ref{sec:Validation}, VERT is validated with two sample problems, and the results are presented in Section \ref{sec:Results}.
Further discussion on the implications and suggestions for practical implementation of VERT is in Section \ref{sec:Discussion}.

\begin{figure*}
 \includegraphics[width=\linewidth]{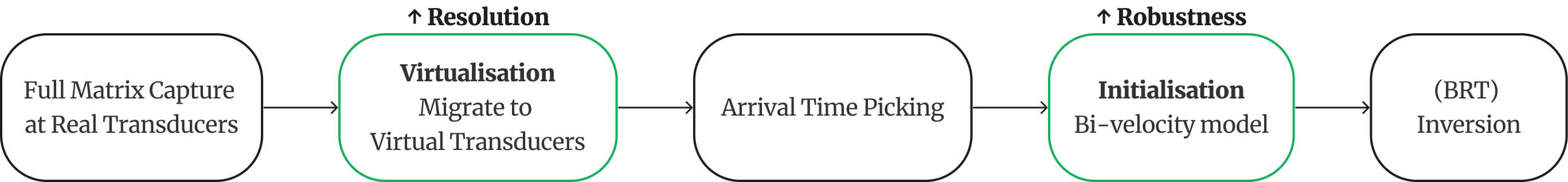}
 \caption{The steps of VERT. Novel steps are shown in green; Classical BRT steps are shown in black.}
 \label{fig:flowchart}
\end{figure*}

\section{Bent-Ray Tomography} \label{sec:BRTVERT}
Before explaining VERT in detail, we will briefly discuss Bent-Ray Tomography (BRT) which VERT improves upon. The wave equation for a scalar property $p$ (e.g. pressure) in the frequency domain is described by
\begin{equation}
 (\nabla^2+k^2)p = 0,
 \label{eq:helmholtz}
\end{equation}
which is the Helmholtz equation, with $k = \omega/c_{phase}$ being the local wavenumber expressed as $\omega$ the angular frequency and $c_{phase}$ is the phase velocity. BRT utilises ray theory, which solves \eqref{eq:helmholtz} by taking the infinite frequency assumption ($\omega \rightarrow \infty$), valid when perturbations in the field are large compared to the wavelength. This leads to the eikonal equation
\begin{equation}
 \frac{1}{c_{phase}} = |\nabla \tau|,
 \label{eq:eikonal}
\end{equation}
where $\tau$ is the time delay for the specific frequency component to travel from a source to any point in the field \parencite{fomel1997traveltime}. In BRT, typically the wavepacket arrival times $\tau_{group}$ are picked, so an alternative form of $1/c_{group} = |\nabla \tau_{group}|$ is used in practice, and we will use $c$ to refer to both velocities generally from now on.

The forward problem in BRT is solved by eikonal solvers such as the fast marching method (FMM), which can calculate $\tau$ for each physical source ($\rSrc$) receiver ($\rRcv$) pair robustly and quickly given the guess velocity field $c$ \parencite{eikonal}. FMM uses a finite difference approximation and represents the wavefront as multiple secondary sources, modelled as binary state points, which tag the nearby points according to the local velocity $c$ \parencite{eikonal}.

The inverse problem in BRT is posed as a minimisation problem in the form of 
\begin{equation} \label{eq:brtMinimise}
 \text{Minimise}~\misfit(\sos{\model}) = \sum_{i,l}\left(\tMat{\model}[i,l](\sos{\model})-\tMat{\meas}[i,l]\right)^2,
\end{equation}
where $\misfit$ is the misfit, $\tMat{\model}[i,l](\sos{\model})$ and $\tMat{\meas}[i,l]$ are respectively the modelled and the experimentally measured arrival times between the $i$-th source and the $l$-th receiver. This inversion step is usually solved through an iterative conjugate-gradient method based on back-tracing the misfit onto the field and updating the corresponding voxels through which they pass \parencite{LiFMM,Hormati2010}.

The eikonal equation \pareneqref{eq:eikonal} only uses the local velocity $c$ in calculating wavefront propagation, showing that travel time is only sensitive to perturbations directly on the ray path in ray theory. The inability of the secondary sources in FMM to interfere with each other also shows that diffraction effects are not modelled. When diffraction effects are properly considered, travel time of a finite frequency wave is sensitive to perturbations in the first Fresnel zone ($F_1$, which is shaped like a banana-doughnut and is widest at the midpoint) which excludes the ray path \parencite{BananaDoughnutHealing}. This discrepancy is behind the effect known as wavefront healing, and introduces an error whereby ray theory is unable to detect perturbations below size \parencite{BananaDoughnutHealing}
\begin{equation}
\resLim{ray} \approx F_{1,max} = \sqrt{ \lambda_0\cdot L } = \sqrt{\frac{c_0\cdot L}{f}},
\label{eq:rayError}
\end{equation}
whereby $f$ is the central frequency of the propagated wave, $L$ is the distance between the source and receiver pair and $c_0$ and $\lambda_0$ are the background sound speed and wavelength respectively. 

BRT's resolution is limited by this error in ray theory which carries to the inversion step. The global optimum from solving \eqref{eq:brtMinimise} can only achieve a maximum resolution proportional to the error intrinsic in ray theory $\resLim{ray}$, giving BRT's resolution limit as
\begin{equation}
\resLim{BRT} \propto \sqrt{\frac{c_0 \cdot \underset{i,l}{\max} |{\rSrc_i - \rRcv_l}| }{f}},
\label{eq:brtLim}
\end{equation}
assuming that the longest ray path is a straight line \parencite{Williamson1991,Williamson1993}. $\rSrc_i$ and $\rRcv_l$ are the positions of the $i^{th}$ source and $l^{th}$ receiver. For circular arrays, $\underset{i,l}{\max} |{\rSrc_i - \rRcv_l}|$ simplifies to the array diameter $D$. The challenge in bone imaging and other extended-range tomography problems is that $c_0$, $f$ and $D$ are intrinsic to the problem, while increasing $f$ usually leads to increased attenuation and scattering, making it difficult for BRT to achieve high resolution in these contexts. 

BRT's robustness is partially augmented by its inherent low resolution, as the back-traced misfit field is usually relatively smooth and monotonic and has few local minima \parencite{Huthwaite.2014}. 
However, with high-contrast problems, a large misfit is back-traced and forms many valleys, increasing the chance of convergence at local minima during gradient descent. 
This mis-convergence sometimes manifests as ray-like or clustered striated artefacts as rays become unable to escape the valley, such that only voxels directly on the ray path are updated.
This effect could be further exacerbated by low sampling. 
Regularisation techniques, like total variation and first/second derivative minimisation, could facilitate convergence towards the global minimum but may limit resolution (e.g. from blurring) and require tweaking. Hence, with increased contrast, the robustness of BRT decreases.

\section{New Technique} \label{sec:New Technique}
VERT addresses BRT's challenges in solving high-contrast extended-range tomography (introduced in Section \ref{sec:BRTVERT}) with two novel steps: virtualisation, which predominantly increases resolution and initialisation, which increases robustness. Figure \ref{fig:flowchart} shows a flowchart of the VERT algorithm, with the novel VERT steps in green and the classical BRT steps in black. We will now discuss virtualisation in Section \ref{sec:Virtualisation} and initialisation in Section \ref{sec:Initialisation}.

\subsection{Virtualisation} 
\label{sec:Virtualisation}
\subsubsection{Motivation}
\begin{figure}[htb] 
 \centering
 \includegraphics[width=\columnwidth]{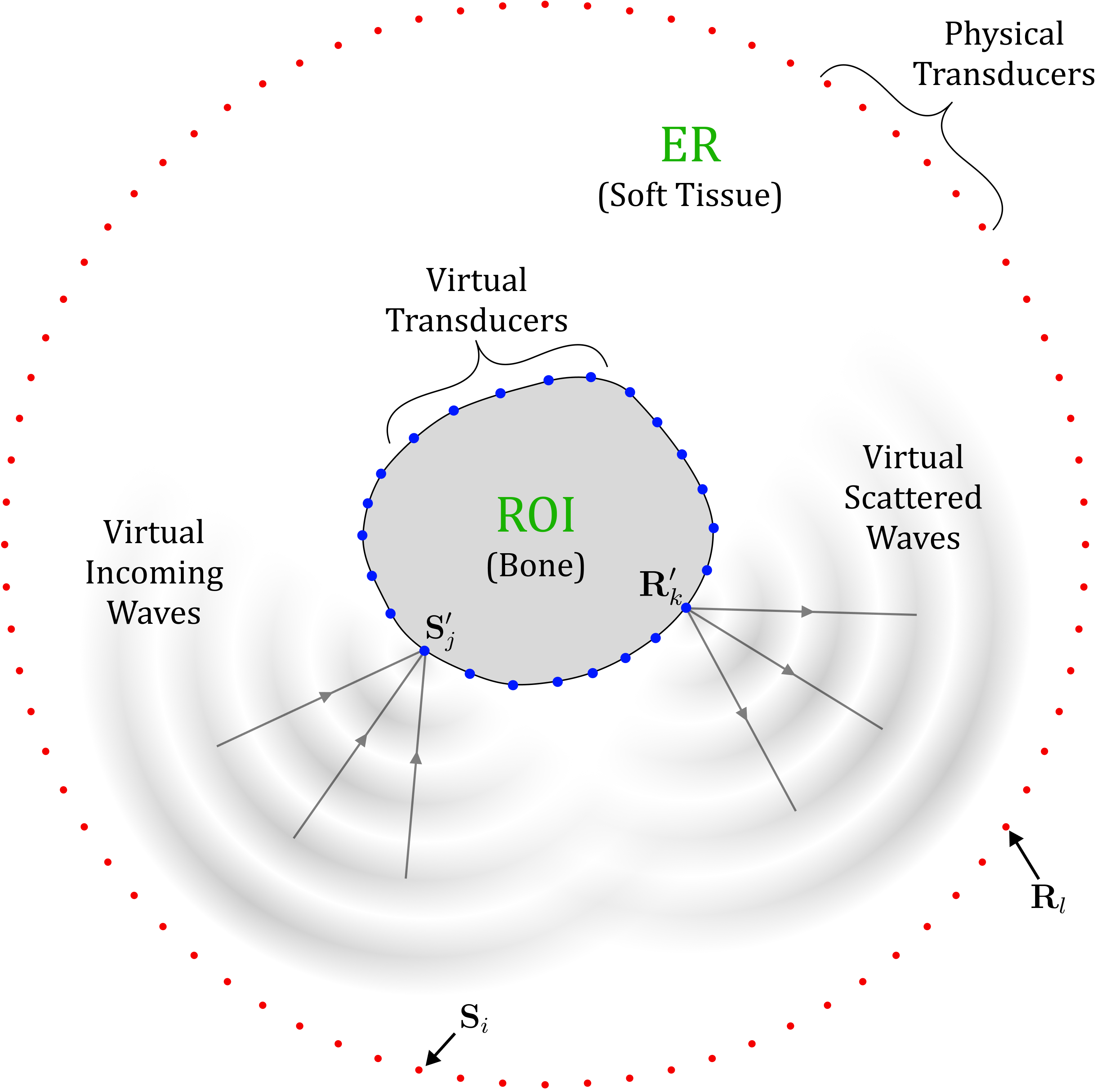}
 \caption{Visualisation of the VERT virtualisation step on a bone imaging setup. The ER (soft tissue) prevents placing physical transducers (red) directly on the ROI; VERT improves the resolution limit by performing BRT from virtual transducers (blue) on the ROI boundary.}
\label{fig:bonesetup}
\end{figure}

The key insight behind VERT is that a flexible, robust and problem-agnostic algorithm like BRT is only required within the unknown (and high-contrast) ROI, as shown in \figref{fig:bonesetup}. 
Conversely, the ER is often quite homogeneous---consisting of various soft tissues in bone imaging, and homogenous waveguide in guided wave tomography. Provided that we have suitable \textit{a priori} information about the low-contrast ER, we can virtualise FMC data from the physical array to the virtual array at the ROI boundary. This boundary corresponds to the minimum convex bounds for concave ROIs or multiple closely clustered ROIs (e.g. corrosion patches). 

Virtualisation brings three main benefits. 
Firstly, performing BRT from the virtual transducers gives VERT's resolution limit as 
\begin{equation} \label{eq:newLim}
\resLim{\text{VERT}}
\propto \sqrt{\frac{c_{\roi}\cdot\underset{j,k}{\max} |{{\vSrc_{j}} - {\vRcv_{k}}}| }{f}}
= \sqrt{\frac{c_\roi \cdot D_{\roi}}{f}} ,
\end{equation}
as calculated from the BRT resolution limit \eqref{eq:brtLim}. $c_\roi$ is the sound speed within the region of interest, $\vSrc_{j}$ and $\vRcv_{k}$ are the positions of the $j^{th}$ virtual source and $k^{th}$ virtual receiver, and $D_\roi$ is the maximum diameter of the region of interest. As $D_\roi$ is always smaller than array diameter $D$, the resolution is improved in VERT.

Secondly, performing BRT from the smaller array of $D_\roi$ drastically increases the speed of the eikonal solver when running at the same resolution. 

Thirdly, virtualisation increases robustness in the case of high-contrast problems as in bone imaging, through a combination of
\begin{enumerate}
 \item unlocking the fast and simple initialisation step (discussed in Section \ref{sec:Initialisation})
 \item separating the regions of high (bone) and low (soft tissue) acoustic impedance, reducing contrast within the region that BRT has to solve (by choosing the ROI boundary strategically)
 \item introducing theoretically unlimited virtual transducers, increasing sampling density.
\end{enumerate}

\subsubsection{Method}
Huygens modelled wavefronts as multiple secondary sources, specifically as monopoles, which \cite{HuygensDipole} corrected to be spatiotemporal dipoles. In a homogeneous background, the spatiotemporal dipole's single-harmonic 3D field is
\begin{equation}
 \phi_\text{StD,3D} \propto X_\text{StD} \frac{e^{ikr}}{r} \left(-ik(1+\cos{\theta})+\frac{\cos\theta}{r} \right),
 \label{eq:spatiotemporal}
\end{equation}
\noindent where $X_\text{StD}=a_\text{StD} e^{-i\omega t}$ is the complex frequency component of the spatiotemporal dipole, $k$ is the wavenumber, $r$ is the radius from the source and $\theta$ is the angle compared to the propagation direction of the spatiotemporal dipole. Thus the 3D spatiotemporal dipole's wavefield fluctuates with $e^{ikr}$, decays with $1/r$ and has an angular dependency. 

In VERT, the goal is for the virtual transducers to mimic these realistic secondary sources (spatiotemporal dipoles) to act as conduits between the ROI and the ER. This would allow virtual excitation and reception at $\vSrc_j$ and $\vRcv_k$ respectively (see \figref{fig:bonesetup}).

\begin{figure}[htbp]
 \centering
 \includegraphics[width=\columnwidth]{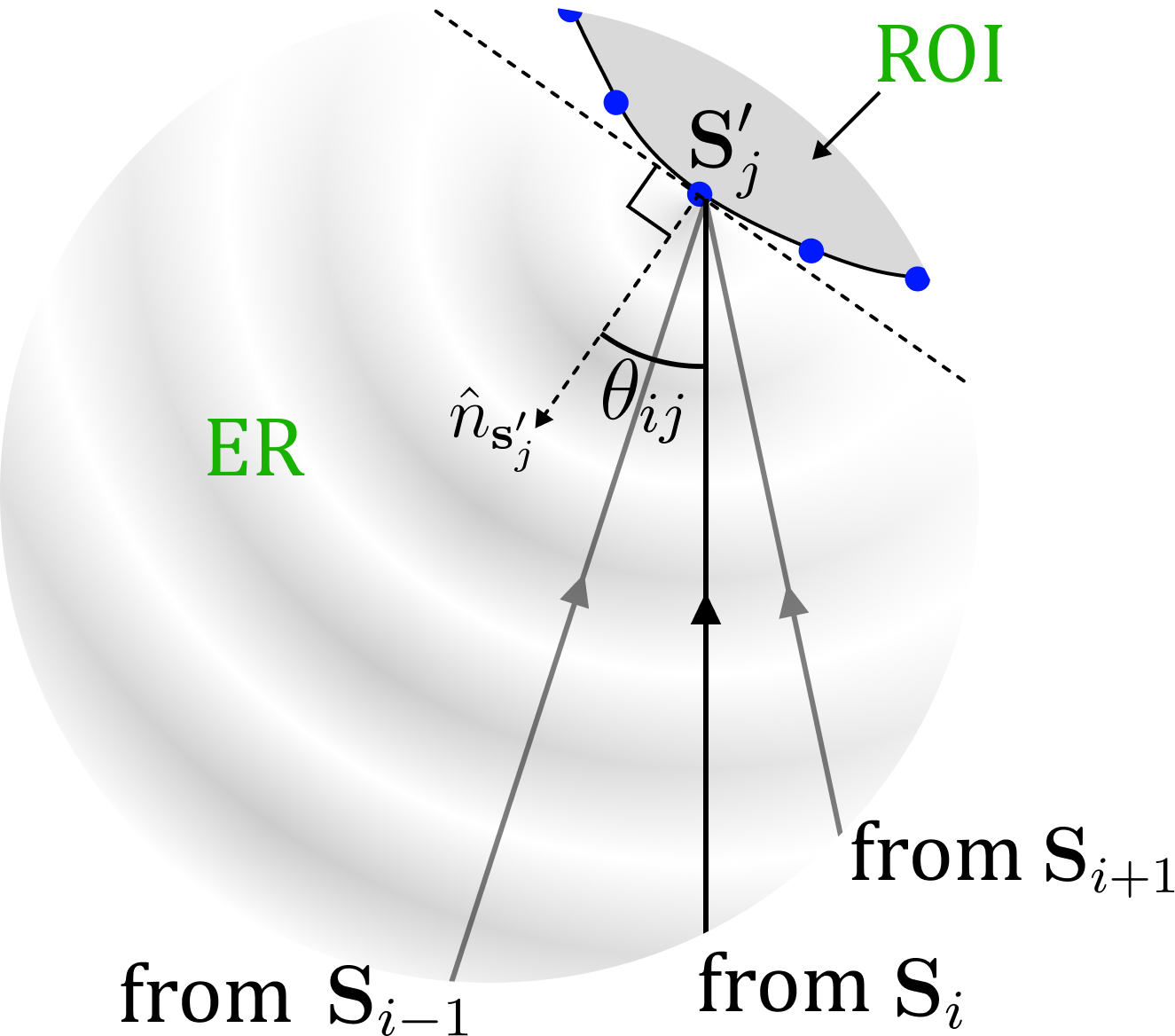}
 \caption{Virtual excitation: incoming waves focused at the $j$-th virtual source, carried out in post-processing}
 \label{fig:VirtualExcite}
\end{figure}

In the physical FMC dataset, each physical transducer sends the same signal which we desire to send from the virtual array. This signal is normally a Hann toneburst which can be taken as a pulse, at $t=0$. The portion of the wavefront relevant to the virtual FMC dataset propagates through the ER to the ROI boundary, interacts with the ROI, and propagates through the ER again to all physical receivers $\rRcv_l$. In the far field, the incoming wavefront from $\rSrc_i$ arriving at $\vSrc_j$($\rSrc_i \rightarrow \vSrc_j$) results in a complex frequency component at $\vSrc_j$ in a homogeneous background of
\begin{equation}
 X_{\rSrc_i \rightarrow \vSrc_j} =
 g(k)
 X_i 
 e^{ikr_{ij}}
 \frac{1}{A\left(r_{ij}\right)}, 
 \label{eq:PlaneWaveDecay}
\end{equation}
where $X_{a\rightarrow b}$ denotes the contribution of signal a on signal b. $g(k)$ is a complex constant that represents multiple factors within propagation, including the overall behaviour of the data acquisition system (e.g. transducers). $X_i=a_i e^{-i\omega t}$ is the complex frequency component of the wave leaving $\rSrc_i$, $r_{ij}=|\vSrc_j-\rSrc_i|$ and $\frac{1}{A}$ is the decay with distance from geometric spreading ($A_\text{3D}=r_{ij}, A_\text{2D} = \sqrt{r_{ij}}$). $e^{ikr_{ij}}$ accounts for the phase shift from the journey $\rSrc_i\rightarrow\vSrc_j$, which is very similar to $e^{ikr}$ of \eqref{eq:spatiotemporal}. This similarity suggests that a wavefield similar to a spatiotemporal dipole can be recreated with sufficient focused wavefronts in the manner of beamforming/delay-and-sum, as shown in \figref{fig:VirtualExcite}.

We convert the physical FMC data to the virtual FMC data by the virtualisation step which defines the virtual signal (denoted as $X'$) at $\vSrc_j$ from incoming waves
\begin{equation}
 X'_j = \sum_i{
 X_{\rSrc_i\rightarrow \vSrc_j}
 e^{-ikr_{ij}} 
 A\left(r_{ij}\right)
 \frac{\theta_{(i+1)j}-\theta_{(i-1)j}}{2}
 B\left(\boldsymbol{\theta}_{ij}\right).
 }
 \label{eq:virtualexcite}
\end{equation}
$e^{-ikr_{ij}}$ and  $A\left(r_{ij}\right)$ are the phase shift and distance weighting factors which undo the corresponding effects of ER propagation on the journey $\rSrc_i\rightarrow\vSrc_j$ described in \eqref{eq:PlaneWaveDecay}. 
In the time domain this phase shift corresponds to a time advance of $\timeshift_{ij}=\frac{r_{ij}}{c_{ER}} = \frac{kr_{ij}}{\omega}$ in a homogeneous ER with speed $\cout$.
This shift normalises and focuses each incoming wave to give the virtual signal $X'_j$ the same phase as the physical signal $X$, similar to delay-and-sum/beamforming. 
$\frac{\theta_{(i+1)j}-\theta_{(i-1)j}}{2}$ accounts for any uneven angular spacing between incoming waves from physical sources $\rSrc_i$ according to the trapezium rule as part of discrete integration.

The directivity pattern $B$ varies with the angle $\theta_{ij}$ defined by the normal $\hat{n}_{\vSrc_j}$ of the ROI boundary at $\vSrc_j$ as shown in \figref{fig:VirtualExcite}. To mimic a spatiotemporal dipole \pareneqref{eq:spatiotemporal} in the far field ($r\gg0$), 
\begin{equation}
 B_\text{StD,far} \approx 1+cos(\theta_{ij})
\end{equation}
gives a cardioid directivity pattern. However, we note that any incoming waves arriving from above
the local tangent (i.e., $|\theta|>\frac{\pi}{2}$) could interact with the unknown and potentially high-contrast ROI, breaking Born's approximation. 

A well-apodized approximation to a cardioid directivity pattern is 
\begin{equation}
B = \max\left( \cos{\theta_{ij}} ,0 \right)^n,
\end{equation}
which is the one-sided directivity pattern of a dipole source when $n=1$. 
This approximation restricts illumination of the virtual source $\vSrc_j$ to the known and no/low-contrast ER, maintaining Born's approximation. We have thus obtained the relationship between all physical sources and all virtual sources.

While this analysis has been performed for a virtual source transducer, on reception the same concept can be applied through the principle of reciprocity, enabling signals to be effectively measured from a virtual transducer in the medium. Instead of wave amplitude being radiated into the medium from a source, the equations would now derive the sensitivity pattern of the transducer within the medium. Thus, the overall relationship between the physical FMC data set ($\rSrc_i \rightarrow \rRcv_l$) and the virtual FMC data set ($\vSrc_j \rightarrow \vRcv_k$) is:
\begin{equation}
\begin{split}
     X'_{\vSrc_j\rightarrow \vRcv_k} &=  \\
     \sum_{i,l}X_{\rSrc_i\rightarrow \rRcv_l}
     &e^{-ikr_{ij}} 
     A\left(r_{ij}\right)
     \frac{\theta_{(i+1)j}-\theta_{(i-1)j}}{2}
      B\left(\boldsymbol{\theta}_{ij}\right)\\
     &e^{-ikr_{lk}} 
     A\left(r_{lk}\right)
     \frac{\theta_{(l+1)k}-\theta_{(l-1)k}}{2}
    B\left(\boldsymbol{\theta}_{lk}\right),
 \end{split}
 \label{eq:Virtualisation}
\end{equation}
which is easily executable as a frequency-wise matrix multiplication. By taking the inverse Fourier transform of $X'_{\vSrc_j\rightarrow \vRcv_k}$ to return to the time domain, we can recover the virtual FMC data from the physical FMC data.

The first point to note is that we have not considered any aberration which would occur if the ER was inhomogeneous. As with many focusing approaches, provided that the ER is known or can be approximated in some way, it should be possible to correct for this. This will not be evaluated in this paper.

A second point to note is that the virtual migration process of FMC data could also be carried out through delay-and-sum focusing in real-time. However, real-time focusing may result in longer data collection times for two reasons. Firstly, any inaccuracies in the \textit{a priori} assumptions (e.g. $c_0$, ROI boundary location) for the focusing law may require re-capturing. Secondly, separate real-time data capture and processing is required to obtain \textit{a priori} information critical to virtualisation. We will not discuss this further in the present paper.

A third point to note is that the virtual focusing step allows placing unlimited transducers on the ROI boundary. This increases spatial sampling within the ROI boundary, which in turn increases the density of ray paths from which inversion occurs. As the distance between rays decrease, robustness is increased as rays are less likely to become stuck in local optima.

\subsection{High-Contrast Initialisation} \label{sec:Initialisation}
As discussed in Section \ref{sec:BRTVERT}, BRT's robustness decreases under high contrast as a large travel time misfit $\misfit$ forms many valleys in the back-traced field, resulting in the conjugate-gradient method converging at local optima. A physical example of this is the reconstruction of a very low-speed object, where, under the eikonal approximation, the fastest ray paths pass either side of the object rather than through it, which means that the reconstructed wave speed within the object will be unconnected to the true value. It is common to initialise the entire domain (ER \& ROI) with the sound speed of the ER ($c_{ER}$), but this only works for low-contrast problems.

Observing that there is usually a clear velocity contrast between the ROI and the ER, we can take the bi-velocity model $\sos{\text{bi-velocity}}$ as a first approximation of the true sound speed map: 
\begin{equation}
 c(\relLoc{}) \approx \sos{\text{bi-velocity}} (\cin,\cout) = 
 \begin{cases}
 \cin & \text{if } \relLoc \in \roi \\
 \cout & \text{otherwise,}
 \end{cases} 
\end{equation}
where $\cout$ is the speed of the ER, and $\cin$ is the unknown initial sound speed within the ROI. \textit{A priori} information on $\cout$ allows us to further simplify to $\sos{\text{bi-velocity}} (\cin,\cout) = \sos{\text{bi-velocity}} (\cin)$

For BRT, initialisation with the bi-velocity starting model is obtained by solving
\begin{multline} \label{eq:brtInit}
 \text{Minimise}~ \misfit(\sos{\text{bi-velocity}}(\cin)) \\
 = \sum_{i,l}
 \left(
 \tMat{\model}[i,l](\sos{\text{bi-velocity}})
 -\tMat{\meas}[i,l]
 \right)^2,
\end{multline}
which is a more robust version of \eqref{eq:brtMinimise}. The misfit and hence arrival time $\tMat{\model}[i,l](\sos{\text{bi-velocity}})$ must be re-calculated through an eikonal solver (e.g. FMM) for each iteration in any optimisation algorithm. These initialisation iterations hence take as long as BRT inversion iterations, making initialisation slow and cumbersome. 

For VERT, initialisation with the bi-velocity starting model is instead obtained by solving
\begin{multline} \label{eq:vertInit}
 \text{Minimise}~ \misfit'(\sos{\text{bi-velocity}}(\cin)) \\= \sum_{j,k}
 \left(
 \tau'^{\,\model}_{j,k}(\sos{\text{bi-velocity}})
 -\tau'^{\,\meas}_{j,k}
 \right)^2,
\end{multline}
where $\misfit'$ is the misfit function for virtual FMC data and $\tau'^{\,\model}_{j,k}$ is the modelled arrival time between $\vSrc_j$ and $\vRcv_k$. $\tau'^{\,\meas}_{j,k}$ is the measured travel time between $\vSrc_j$ and $\vRcv_k$. Crucially, this virtual measurement hinges on the virtualisation step discussed in Section \ref{sec:Virtualisation}.
Given convexity of the ROI boundary, it is trivial to calculate the arrival times $\tau'^{\,\model}_{j,k}(\sos{\text{bi-velocity}}) = \frac{||\vRcv-\vSrc||}{\cin}$ and hence the misfit function. The optimisation in \eqref{eq:vertInit} can thus be solved with a golden section search (or other rudimentary algorithms) to rapidly iterate towards $\cin$, giving the bi-velocity starting model. 

\begin{figure*}
 \centering
 \adjustbox{
 trim = 0.5cm 0cm 0cm 0cm,
 clip,width=\linewidth}{\includegraphics{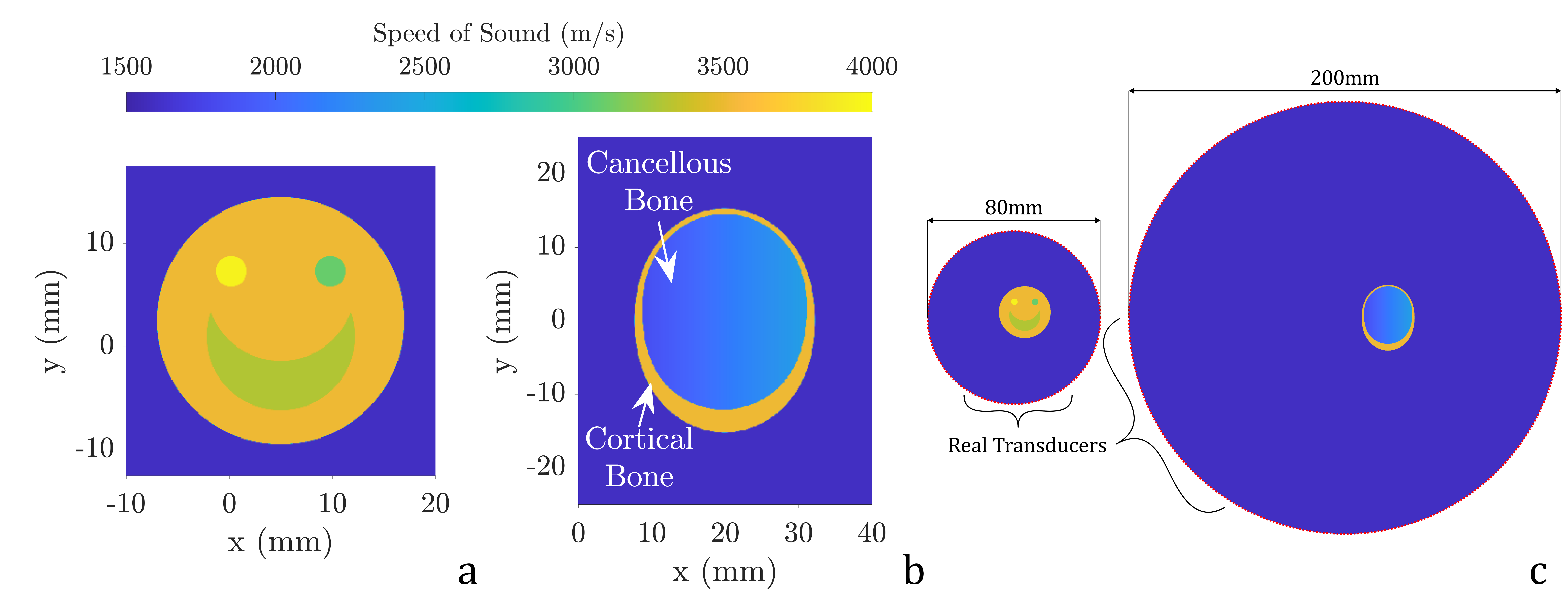}}
 \caption{(a) Smiley80 detailed view (b) Bone200 detailed view (c) Both problems shown to scale within the physical array (transducers in red)}
\label{fig:SampleProblems}
\end{figure*}

VERT's high-contrast robustness is hence increased by this rapid and simple initialisation step. By reducing the initial travel time misfit, the initialisation step reduces the number and depth of fluctuations in the back-traced misfit field despite high contrast, reducing local minima and increasing robustness. Similar to Section \ref{sec:Virtualisation}, this only requires \textit{a priori} information about the exterior but not the high-contrast unknown interior.

\section{Validation Methodology} \label{sec:Validation}
Given the novel nature of the proposed algorithm, we have distilled the problem into two key challenges which VERT needs to overcome to be successful:
\begin{enumerate}
 \item Inaccessibility of the region of interest 
 \item High contrast between the region of interest and the surroundings 
\end{enumerate}
If the two challenges are unresolved, the first results in a limited resolution, and the second in the BRT step converging to a local, rather than global, optimum. Our primary goal is to show that VERT achieves increased resolution and robustness compared to BRT. As secondary goals, we want to characterise VERT's performance against the theoretical resolution limit given in \eqref{eq:newLim}, and establish the potential merits of VERT in bone imaging. 

In this vein, we will prioritise characterising the algorithm's performance against these metrics, rather than the realism of the sample problems posed (Section \ref{sec:Sample Problems}). We have validated VERT \textit{in silico} and have limited ourselves to a 2D problem, which is discussed in Section \ref{sec:FMC}. We have also not carried out fine-tuning of variables, such as frequency and signal choice, which are largely constrained by the problem at hand. Section \ref{Sec:Post-Processing} discusses the practical implementation of VERT and other compared scenarios. Section \ref{sec:Results} presents the reconstructions and discusses our findings.

\subsection{Compared Imaging Configurations}
We have chosen 4 different imaging configurations in addition to the input model (GroundTruth) for comparison, and these configurations will be labelled as VERT, BRT, BestBRT and NearBRT, of which VERT and BRT have already been discussed in detail in this paper.

BestBRT represents a realistic best-case scenario for BRT to demonstrate that any resolution gains in VERT are not purely from the supplied \textit{a priori} information ($C_{ER}$ and ROI boundary). An initialisation step is added before the inversion step in \figref{fig:flowchart} which solves the more cumbersome \eqref{eq:brtInit} to obtain a starting bi-velocity model. The same FMC data as VERT and BRT is used. 

NearBRT represents the inherent resolution limit of BRT at the normally inaccessible boundary, by artificially introducing transducers (dipoles) \textit{in silico} at the same locations as virtual transductabers in VERT. It is the only configuration that does not use the FMC data at the physical array. Initialisation with NearBRT the bi-velocity model is carried out by solving the simple \eqref{eq:vertInit} before the inversion step.

\subsection{Sample Problems} \label{sec:Sample Problems}
The setup for the 2 sample problems largely resembles that shown in \figref{fig:bonesetup} with the region of interest being substituted according to the specific problem. Although this is of secondary concern in evaluating VERT's efficacy, we have endeavoured where we can to use realistic values (e.g. speed of sound, density) for biological tissues from \cite{ITIS}, the most relevant of which are highlighted in Table \ref{tab:ITIS}.

\begin{table}
 \centering
 \caption{Properties of biological tissues \parencite{ITIS}}
  \label{tab:ITIS}
 \begin{tabular}{ccc}
 \hline\hline
 Tissue& Velocity$^a$ (ms$^{-1}$)&Density (kg m$^{-3}$) \\
 \hline
 Cortical Bone& 3514.9 $\pm$420.3 &1908 $\pm$ 133\\
 Cancellous Bone& 2117.5$\pm$288.7 &1178 $\pm$ 149\\
 Soft Tissue$^b$& 1588.4$\pm$21.6 &1090 $\pm$ 52\\
 \hline\hline
 \multicolumn{3}{p{0.9\columnwidth}}{ $\pm$ denoting the standard deviation (s.d.) of the properties.}\\
    \multicolumn{3}{p{0.9\columnwidth}}{$^a$Velocity corresponds to dilational wave velocity.}\\
 \multicolumn{3}{p{0.9\columnwidth}}{$^b$Muscle is used as the proxy value for soft tissue in the ER for both sample problems.}
 \end{tabular}
\end{table}

\subsubsection{Smiley Problem (Smiley80)}
The first sample problem is that of a 2D bone smiley face embedded within soft tissue (Smiley80), shown in \figref{fig:SampleProblems}(a). This sample problem was chosen as an initial test to aid visualisation of the resolution gains from using VERT with easily reproducible and analysable geometric shapes. \figref{fig:SampleProblems}(c) shows the full setup where the ROI is offset by $x=+5mm, y=+2.5mm$ relative to the centre of the 80mm diameter array. 

The detailed geometry and material properties (with reference to Table \ref{tab:ITIS}) are as follows. The diameter of the circular smiley face is 24mm, where the properties of cortical bone are used on the face of the smiley, and that of muscle is used for the surrounding soft tissue. The eyes are located 4.8mm above the centre of the smiley and are spaced 9.6mm apart from each other. The mouth is located 1.5mm below the centre of the face, and is mathematically described by a circle 14.4mm in diameter where the points above the line described by 2.4mm above its centre are reflected across the line. The left eye, right eye and mouth have velocity and density 1 s.d. above, 1 s.d. below and 0.5 s.d. below that of cortical bone respectively. 

\subsubsection{Simplified Bone Model (Bone200)}
The second sample problem (Bone200) uses reproducible geometries and simple material properties to aid the characterisation of VERT's performance in femoral neck imaging.
\figref{fig:SampleProblems}(b) shows a detailed view of the ROI (a simplified ellipsoid bone model), with cancellous bone encased by cortical bone. \figref{fig:SampleProblems}(c) shows the full setup, where the ROI is surrounded by soft tissue (the ER) as would be the case \textit{in vivo}. 

We based Bone200 with female reference data as females are more likely to suffer from osteoporosis. 
The asymmetrical periosteum (outer cortical surface) is simplified as an ellipsoid (minor axis 24.5 mm, major axis 30.5 mm; minor/major ratio: 0.8), with the cortical thickness distributed as the average between fracture cases and age-matched controls \parencite{zebaze2005} \cite{Bell1999}. 
The periosteum and endosteum (inner cortical surface) are then constructed with cubic spline interpolation of the values provided in Table \ref{tab:spline} with periodic end conditions. 
\begin{table}[h]
 \centering
 \caption{Values used for construction of the Bone200}
 \label{tab:spline}
 \begin{tabular}{ccc}
 \hline\hline
 Azimuth Angle (degrees)& Diameter (mm)& Thickness (mm)\\
 \hline
 0& 24.5& 1.12\\
 45& 27.5& 0.72\\
 90& 30.5& 0.75\\
 135& 27.5& 0.72\\
 180& 24.5& 1.07\\
 225& 27.5& 1.97\\
 270& 30.5& 3.14\\
 315& 27.5& 2.16\\
 \hline\hline
 \end{tabular}
\end{table}

The material properties include soft tissue for ER, homogeneous cortical bone for the entire cortical region, and a varying gradient of cancellous bone velocity and density of -1 s.d. to +1 s.d. horizontally from left to right (refer to Table \ref{tab:ITIS}). This gradient is introduced to differentiate between imaging capabilities for cancellous and cortical bone, as both contribute to bone strength. It is set in an orthogonal direction to the largely vertical change in cortical thickness to separate the effects. For the cancellous and cortical bones, we decided against a more realistic porous structure to simplify the reconstruction comparison. Replacement with empirical values from \cite{ITIS} was possible as the first arrival wave in cancellous bone, the fast Biot wave, is non-dispersive \parencite{BovineBiot}. We have simplified the ER by assuming all soft tissue (muscle, fat) to have the same properties as muscle.

The ER size is of specific concern as this affects the minimum array diameter. \textit{In-vivo} femoral neck imaging must be carried out at an oblique angle to obtain a cross-section along the femoral neck axis and avoid the pelvic bone. This oblique slice starts between the greater trochanter and the iliac crest on one side, and ends below the inferior pubic ramus (sit bone), and has a diameter of approximately 0.7$\times$ buttocks circumference. The ER (soft tissue) diameter, approximated as a circle, would then be 200-300mm for females across ethnicity and age groups from anthropometric data \parencite{Anthropometric}. For the current \textit{in silico} study (\figref{fig:SampleProblems}c), we conservatively use a 200mm diameter for the ER with the transducers spaced equally around its boundary, and introduced a 20mm horizontal offset to the bone to account for its anatomical location. 

\subsection{Full Matrix Capture} \label{sec:FMC}
Full Matrix Capture (FMC) is carried out \textit{in silico} through Pogo \parencite{Pogo}, an FE modelling software run on GPUs. We used the fluid-solid coupling mode using fully-displacement-based models \parencite{Yiannis}, with soft tissue being treated as a fluid and cortical and cancellous bone as solids with Poisson's ratio of 0.3. An unstructured triangular mesh with 30 elements per ER wavelength was used, with a 25\% denser grid at boundaries and transducers. To largely eliminate boundary reflections such that the simulated domain is a subsection of an infinite space, an absorbing boundary using the stiffness reduction method \parencite{SRM} was added to the limits of the circular propagation space. Both the propagation space and absorbing boundary were adjusted to limit the simulation burden, whereby the absorbing boundary thickness is at least 9 wavelengths thick, and the distance between the array and the absorbing boundary is at least 10mm. 

A 5-cycle 1-MHz toneburst signal is sent from transducers. In NearBRT, a single node is inserted at the locations of the VERT virtual transducers. These transducers transmit the signal as a force in the normal direction of the local ROI boundary, corresponding to a dipole dilational wave. In all other cases, the physical transducers shown in red in \figref{fig:SampleProblems} transmit and receive purely dilational waves.

The spacing between physical array transducers is equal to half of the wavelength of the ER (soft tissue) such that the array is sufficiently sampled (317 for Smiley80, 792 for Bone200). For NearBRT, the virtual transducers on the ROI boundary are replaced \textit{in silico} by the physical transducers (300 for Smiley80, 400 for SimpBone200).

Given the varying material properties used and the varying mesh density, a very conservative value is used for the time step ($t = 5.76 ns$) to always satisfy the Courant criterion ($\text{Courant number}\approx 0.4$). Forward simulations of Smiley80 took 25 minutes and Bone200 took 11 hours to run on a computer with an NVIDIA GeForce RTX 2080 Ti (4-core) GPU. We note that FMC \textit{in vivo} or \textit{ex vivo} does not require FE simulation.

\subsection{Post-Processing} \label{Sec:Post-Processing}
The FMC data from Pogo are transferred into MATLAB (R2023b) for further processing on a computer with Intel(R) Xeon(R) CPU E5-2667 v4 @ 3.20GHz (16 core, 32 parallel threads). 

\subsubsection{Virtualisation (VERT only)} 
In the case of VERT, we first need to migrate the FMC time traces from the physical array to the virtual transducers, which we space equally around the ROI boundary (300 for Smiley80, 400 for SimpBone200). These are oversampled since there are no restrictions, and oversampling increases robustness. In preparation for the virtualisation step, a cosine ramp filter is applied to smoothly taper off the last 10\% of the FMC time traces. The remaining time traces are Fourier transformed into the frequency domain before the lowest 1\% of frequencies are filtered out. 

Based on the locations of physical and virtual transducers and the background speed between them, we calculate the corresponding signal correction. The signal correction for both the transmission phase and the receiving phase is applied through a page-wise matrix multiplication where each page corresponds to a single frequency. This is then inverse Fourier transformed back to the time domain to form the virtual FMC time traces. We time window these traces to only retain the much shorter, relevant portion, since the virtual FMC wavefronts only have to traverse the ROI, rather than the entire diameter of the physical array. Overall, the virtualisation step takes 14 seconds for Smiley80 and 132 seconds for Bone200. 

The angular weighting $B$ in \eqref{eq:Virtualisation} was applied with $n=1$ for Smiley80 and $n=2$ for Bone200. This was found to make arrival time picking procedurally easier with no influence on the actual arrival time.

\subsubsection{Arrival time picking (all cases)} 
For VERT, in preparation for arrival time picking, the virtual time traces are high-pass filtered by means of rolling mean removal to remove low-frequency components. We found that low-frequency artefacts can arise from the fluid-solid implementation of Pogo---after the arrival of a wave, energy at the nodes may take longer than expected to dissipate. These artefacts are non-sinusoidal but instead follow a geometric decay pattern, so could not be cancelled out through interference. The sum of hundreds of these artefacts present in the physical FMC data amplifies the artefact in the virtual FMC data to a level that it can mask the arrival wavefront.

For all cases, we then pick arrival times through a combination of algorithmic picking (e.g. zero-crossing-based, thresholding-based, moving-max-based) followed by manual inspection and correction. More sophisticated and fully automated methods exist, but this is not the focus of the present paper. We have excluded this element from processing time calculations.

\begin{figure*} \centering
 \adjustbox{trim = 2.5cm 1cm 2.75cm 2.5cm
 ,clip,width=\textwidth}{\includegraphics{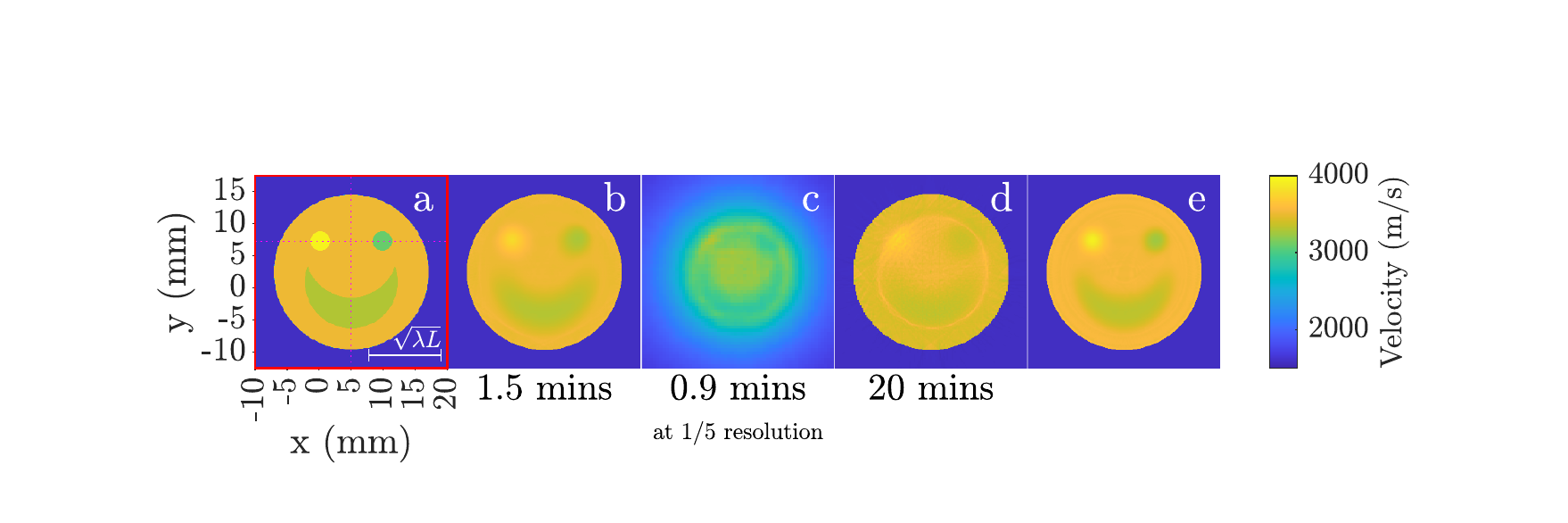}}
 \caption{Smiley80: (a) GroundTruth; (b) VERT; (c) BRT; (d) BestBRT; (e) NearBRT. Processing times are shown below physically implementable algorithms}
\label{fig:SmileIm}
\end{figure*}
\begin{figure*}
 \centering
 \adjustbox{trim = 1.2cm 0.25cm 0.2cm 0cm
 ,clip
 ,width=\textwidth}
 {\includegraphics{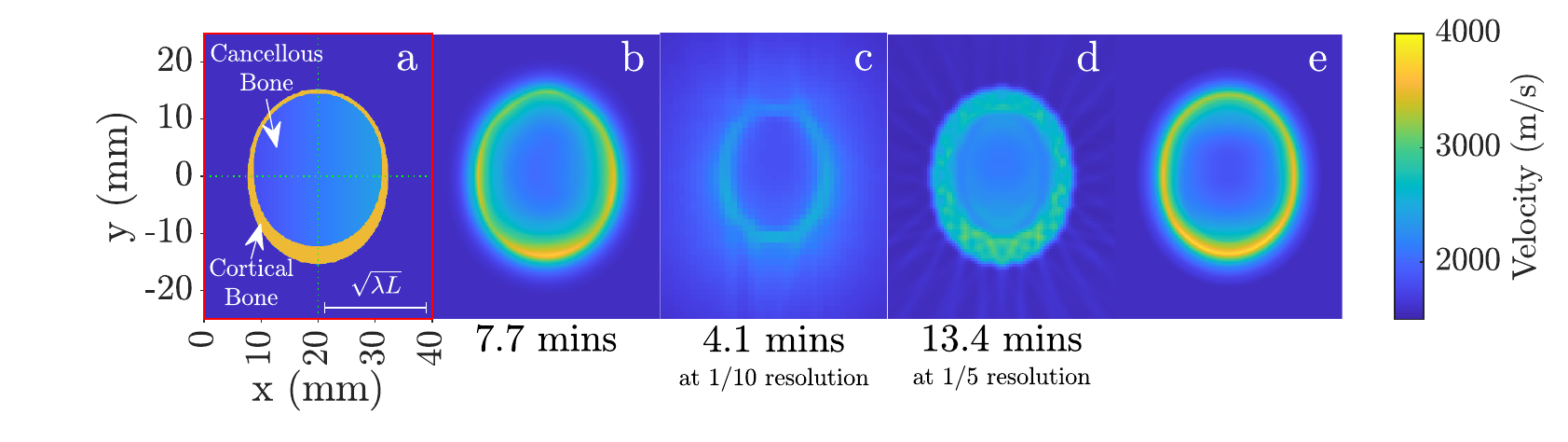}}
 \caption{Bone200:
 (a) GroundTruth; (b) VERT; (c) BRT; (d) BestBRT; (e) NearBRT. Processing times shown below physically implementable algorithms}
 \label{fig:BoneIm}
\end{figure*}

\subsubsection{Initialisation (all cases except BRT)}
Based on the picked arrival times, the background sound speed is initialised to a bi-velocity starting model (see Section \ref{sec:Initialisation}). 

For both NearBRT and VERT, a golden section search between 1500m/s and 4000m/s is used to solve the optimisation problem \pareneqref{eq:vertInit} until $\cin$ converges to within 1m/s, taking 0.03 and 0.04 seconds for Smiley80 and Bone200 respectively.

For BestBRT, the optimisation problem formulated in \eqref{eq:brtInit} was used instead in the golden section search, which requires an eikonal solver. We used the same FMM-based eikonal solver as that in the forward model of the inversion step (Section \ref{sec:inversion}), implemented in C++ \parencite{eikonal}. For Smiley80, this took 6 minutes to compute at 0.1mm resolution, while for Bone200 this took 2.6 minutes at 0.5mm resolution. 

\subsubsection{Inversion (all cases)} \label{sec:inversion}
Finally, for the BRT inversion step, we used an algorithm based on \cite{LiFMM}\cite{Hormati2010} which is implemented in C++. Taking the bi-velocity starting model, the data was reconstructed using a minimal amount of regularisation, which are not fully optimised but consistent (total variation regularisation for both sample problems, and first derivative regularisation is additionally applied on Bone200). 

The default resolution is set at 0.1mm. In some cases, stability issues with the eikonal solver were apparent, and in these cases, or where it is very obvious that the algorithm converged to a local optimum, the inversion step is re-run with a worse resolution until the aforementioned problems are largely resolved. The inversion step took both VERT and NearBRT 1.3 minutes for Smiley80 and 5.5 minutes for Bone200. 

\section{Results} \label{sec:Results}
The reconstructions of Smiley80 and Bone200 are shown in \figref{fig:SmileIm} and \figref{fig:BoneIm} respectively, where lines in each panel (a) mark the locations where cross-sections are extracted, shown in \figref{fig:SmileProf} and \figref{fig:BoneProf}. In Smiley80 the horizontal cross-section is taken at the height of the eyes, whereas all other cross-sections in Smiley80 and Bone200 pass through the centre of the ROI. 

\subsubsection{BRT} \label{sec:BRTVERTResult}
It is clear that VERT far outperforms BRT from \figref{fig:SmileIm}a-c and \figref{fig:BoneIm}a-c. 

The robustness of BRT is far worse than VERT as it was unable to invert at the same resolution. This is exacerbated by the increased array diameter in Bone200. Despite requiring even lower resolution for inversion to be carried out successfully, \figref{fig:BoneIm}(b) still shows ray-like artefacts that signify low robustness. Since the first arrival wave consists largely of direct transmissions between physical transducers, there are very few arrival time pairs that encode information about the ROI. This reduces spatial sampling and encourages the convergence to local minima, giving ray-like artefacts.

\begin{figure*}
\centering
\adjustbox{trim = 1.2cm 0.25cm 2.2cm 0.5cm
 ,clip
 ,width=\textwidth}
{\includegraphics{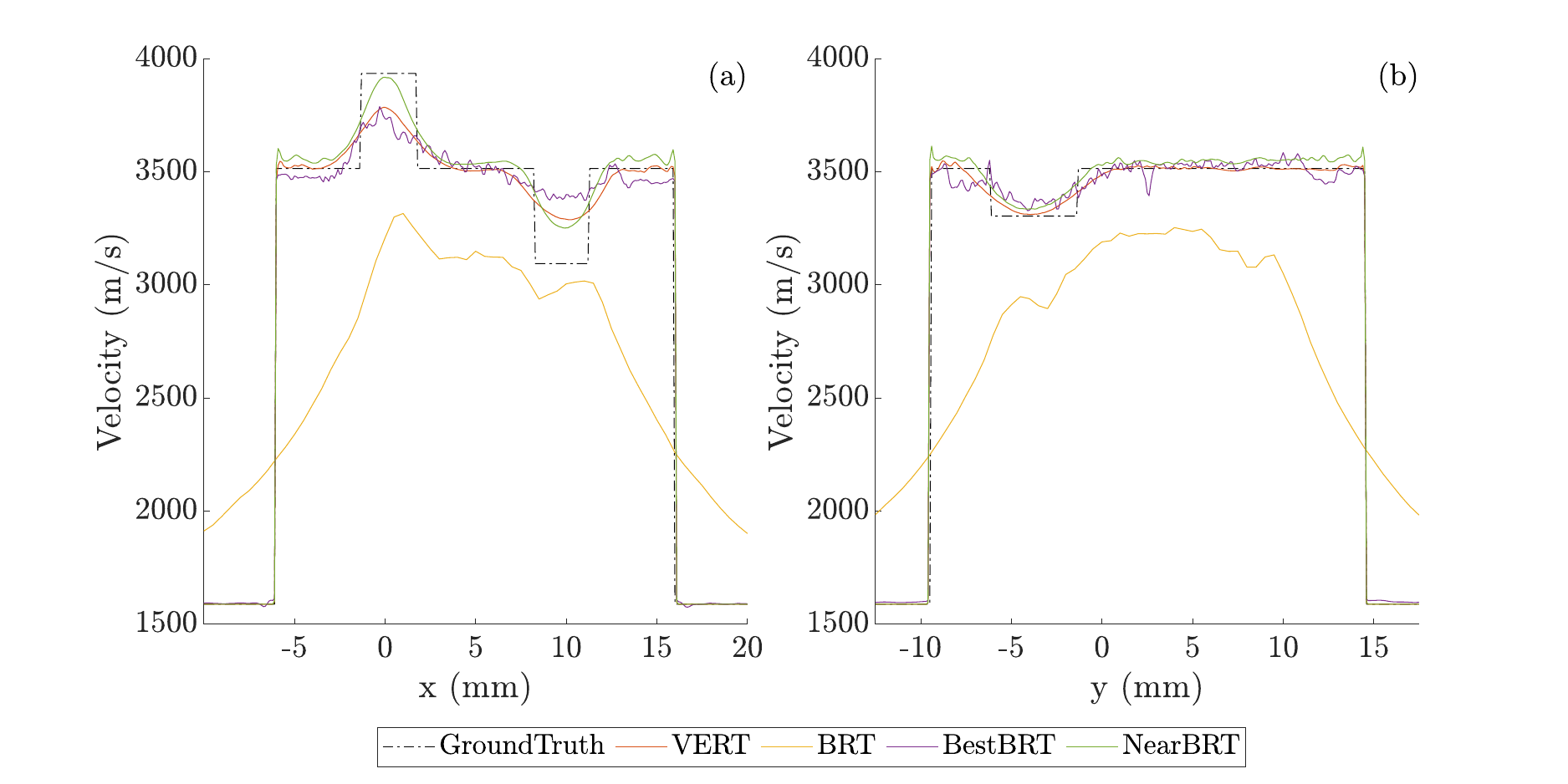}}
\caption{Cross-sections of Smiley80: (a) horizontal at eye height; (b) vertical at the centre.}
\label{fig:SmileProf}
\end{figure*}

\begin{figure*}
\centering
\adjustbox{trim = 1.2cm 0.25cm 2.2cm 0.5cm
 ,clip
 ,width=\textwidth}{
\includegraphics{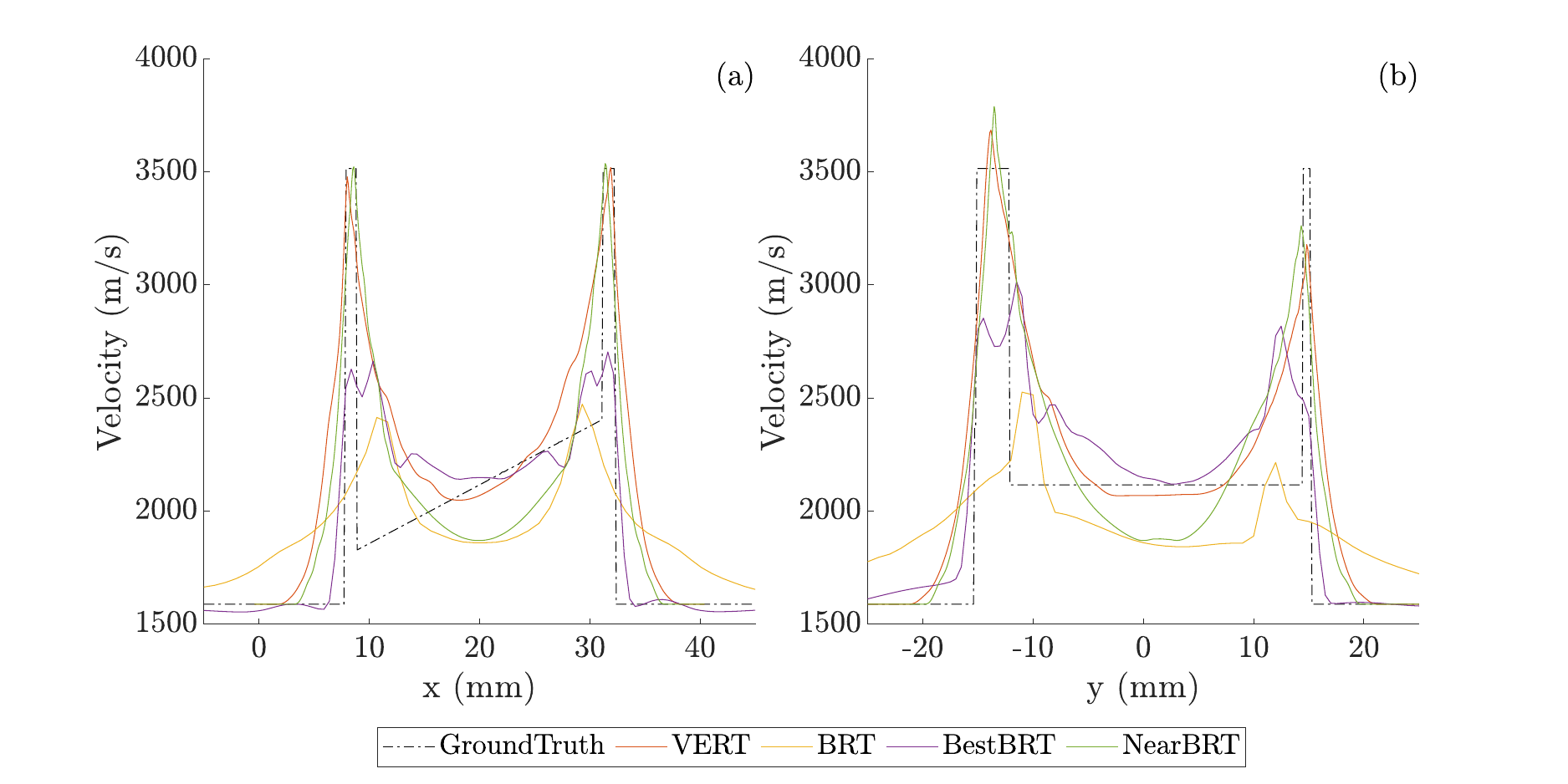}}
\caption{Cross-sections of Bone200: (a) horizontal; (b) vertical.}
\label{fig:BoneProf}
\end{figure*}

In terms of resolution, BRT gives a very poor reconstruction of the bulk geometry, which contains significant morphological inaccuracies. Both of the ROIs in Smiley80 and SimpleBone200 have shrunk in size, by approximately 40\% and 24\% respectively when measured from the first and last local maximum from the cross-sections in \figref{fig:SmileProf}b and \figref{fig:BoneProf}a, although we note that this measurement is affected by the poorly defined ROI boundary, with values of the ER and the ROI leaking into the opposite region. In \figref{fig:SmileIm}(b), we can see that the facial features of Smiley80 are largely fused. Wavefront healing contributes to all of the above problems.

In \figref{fig:BoneIm}(b), no significant variation in cortical thickness or cancellous bone is visible. The cancellous bone is seen to be horizontally symmetric in \figref{fig:BoneProf}a. The cross-section in \figref{fig:BoneProf}b shows some sort of \textit{vertical} internal gradient which does not exist in the GroundTruth, suggesting that the thicker lower cortical bone (which has a uniform speed of sound) actually resulted in BRT incorrectly predicting a higher speed of sound in the cancellous bone.

\subsubsection{BestBRT} \label{sec:BestBRT}
BestBRT, which is supplied with \textit{a priori} information ($C_{ER}$ and ROI boundary) and undergoes the less efficient initialisation step, performs much better than BRT. However, the underlying errors in the ray-based forward model of BRT are unresolved, resulting in a subpar reconstruction in the smaller Smile80, and a highly distorted reconstruction in the larger Bone200. The discrepancies between BRT's result and GroundTruth show that VERT is needed to fully exploit the \textit{a priori} information.

BestBRT robustness is still poor. While Smiley80 could be inverted at the default resolution, the larger Bone200's initialisation and inversion steps were both carried out at a lower resolution. Both of these are better than BRT but worse than VERT. 
In addition, ray-like artefacts are visible in both Smiley80 and Bone200, which encourage convergence to local minima and signify poor robustness. In Smiley80, as shown in \figref{fig:BoneIm}d, they appear as a centred ring-like artefact of diameter $\approx$ 17mm, as well as near the ROI boundary, the centre of the smiley and as a cluster of rays near the left eye. In the larger Bone200, the ray-like artefacts are visible not only inside the ROI boundary but also radiating outwards into the ER (see \figref{fig:BoneIm}d). This is despite the lower resolution and the presence of additional regularisation in the inversion step (Section \ref{sec:inversion}) for Bone200. This suggests that the need for VERT increases with array diameter.

BestBRT resolution is slightly improved in Smiley80 when compared with BRT but still poor in Bone200, which has a larger array diameter.

In Smiley80, we note that the aforementioned ring-like artefact in BestBRT (\figref{fig:SmileIm}d) has a similar shape and diameter to the ROI boundary of BRT's reconstruction (\figref{fig:SmileIm}c). 
This suggests that the error originates from inaccuracies embedded in $\tMat{\model}$ calculated from the ray-based forward model, which results in wavefront healing. This artefact masks the poor resolution of Smiley80. In \figref{fig:SmileProf}a, the height of the left eye's reconstruction is boosted by the ring-like artefact, whereas the right eye only reflects 30\% of the velocity change that exists in the GroundTruth. Both eyes are larger than they should be and have poorly defined boundaries. The ring-like artefact also creates the illusion of a clear delineation between the mouth and the ``chin'' by increasing the velocity sharply at the lower edge, when in reality the BestBRT mouth, as shown in \figref{fig:SmileIm} and \figref{fig:SmileProf}b, is a smudge that extends far beyond the true mouth. 

In Bone200, BestBRT poorly reconstructs both the cortical and cancellous bone regions. The ``cortical bone" is extraordinarily thick with little meaningful variation, with the endosteum (inner cortical boundary) bearing much less resemblance to the endosteum of the GroundTruth (\figref{fig:BoneIm}a) than the location of the overall cortical region in the BRT result (\figref{fig:BoneIm}c). The cross-sections (\figref{fig:BoneProf}) confirm this with the clear alignment between BestBRT and the mislocated BRT peaks. These peaks reflect approximately 50\% of the true velocity variation, compared to the surroundings. The cancellous bone has values closer to GroundTruth than BRT as shown in \figref{fig:BoneProf}, but the horizontal gradient is not reflected by the symmetrical cancellous properties in \figref{fig:BoneIm}d and \figref{fig:BoneProf}a. Similar to BRT but to a lesser degree, BestBRT incorrectly predicts a \textit{vertical} internal gradient in \figref{fig:BoneProf}b, influenced by the change in cortical thickness. This shows that the \textit{a priori} information and the initialisation step alone are unable to correct for the intrinsic errors in the ray-based forward model, especially in extended-range tomography as in bone imaging. 

BestBRT is much slower than both VERT and BRT, taking 13$\times$ the reconstruction time of VERT for Smiley80 while requiring the same \textit{a priori} information. Considering that VERT is faster, higher resolution and more robust, and that BRT does not require \textit{a priori} information and is faster, BestBRT does not have any practical edge in tomography. It does however show that for high-contrast extended-range tomography, the \textit{a priori} information about the ER and the location of the ROI boundary are invaluable.

\subsubsection{VERT and NearBRT} \label{sec:VERT and NearBRT}
The robustness of VERT and NearBRT are much better than BRT and BestBRT. Ray-artefacts are absent in both sample problems (see \figref{fig:SmileIm}b,e and \figref{fig:BoneIm}b,e). In addition, both VERT and NearBRT can be inverted at the default high resolution without problems.

At first glance, VERT and NearBRT are nearly identical and both much better than BRT and BestBRT for both sample problems(\figref{fig:SmileIm}, \figref{fig:BoneIm}). Both are slightly blurred versions of GroundTruth, visible in the eyes and the mouth for Smiley80 (\figref{fig:SmileIm}a,b,e), and around the cortical bone of Bone200 (\figref{fig:BoneIm}a,b,e). This blurring is expected as the size of scatterers approaches the resolution limit evaluated from the ROI boundary \pareneqref{eq:newLim}, and the similarity between VERT and NearBRT shows that VERT is close to achieving this limit. 

For Smiley80, it preliminarily appears from \figref{fig:SmileProf}a that the reconstruction of the left eye is much better in NearBRT than in VERT. However, we note from both NearBRT profiles in \figref{fig:SmileProf} that there seems to be a systematic increase of $\approx$ 25m/s in velocity throughout the reconstructed ROI. This is likely due to a systematic error of picking slightly earlier than the actual arrival times for this case (e.g. due to an overly sensitive threshold). Correcting for this systematic error in \figref{fig:SmileProf}a, we can reinterpret that the reconstruction by NearBRT is equally good for each eye, and in both instances, NearBRT slightly outperforms VERT in accuracy of peak height and width. Similarly in \figref{fig:SmileProf}b, we can see that VERT and NearBRT have nearly identical peak heights for the mouth, both reaching 100\% of the GroundTruth depth. The peak width of the mouth reconstructed by VERT is slightly worse. Closer inspection of \figref{fig:SmileIm}b,d reveals that the features in VERT are indeed slightly blurrier, especially for the smaller features (eyes).

The extra blurring in VERT is likely due to two factors. Firstly, the transducers in NearBRT were modelled as a single node in Pogo, with the signal applied as a force normal to the ROI boundary, giving an unphysical point source transducer. Such transducers cannot be procured, and so NearBRT truly represents a theoretical rather than practical limit. Secondly, any focusing error during the virtualisation step of VERT would result in a larger, finite excitation region on the ROI boundary, equivalent to the averaged sum of several point sources. The combined wavefront from these equivalent sources will arrive at different times to the infinite point source modelled by NearBRT, resulting in a blurring effect. This effect is more apparent for the smaller features where feature boundaries are much closer to each other. This also explains the slightly smoother profile in \figref{fig:SmileProf}b of VERT when compared to NearBRT. 

VERT and NearBRT reconstructions in Bone200 (\figref{fig:BoneIm}) are much more accurate than BRT and BestBRT but appear slightly blurrier than those in Smiley80 (\figref{fig:SmileIm}). 
This is expected as a first derivative smoothing regularisation had to be added due to problems with the inversion step (Section \ref{sec:inversion}), arising from the fact that ray theory cannot model guided waves, which travel circumferentially along the cortical bone. 

Surprisingly, the cancellous interior is much better reproduced by VERT than NearBRT. The vertical profile in \figref{fig:BoneProf}b shows that NearBRT completely overshoots on the cancellous interior value, whereas horizontally (\figref{fig:BoneProf}a) the cancellous region is symmetrical, reflecting none of the gradients that exists in GroundTruth. This arises from the first arrivals of the NearBRT data consisting entirely of the guided circumferential lamb-type (CLT) wave. This results in through-transmission waves, which encode information about the cancellous interior, being excluded from the reconstruction process and the loss of this information. In VERT, the cancellous interior is well-modelled for the central third, closely reflecting the horizontal gradient in the GroundTruth (\figref{fig:BoneProf}a), and the non-sloping nature of the GroundTruth in the vertical profile (\figref{fig:BoneProf}b). This is due to the finite excitation spot in VERT partially suppressing the CLT wave, reducing its interference with the through-transmitted wavefront, which becomes easier to pick as the first arrival.

The cortical reconstructions (\figref{fig:BoneIm}) of GroundTruth by VERT and NearBRT are much more accurate than those from BRT and BestBRT. The horizontal profile (\figref{fig:BoneProf}a) shows that VERT and NearBRT peak heights match GroundTruth almost perfectly for the sides, whereas the vertical profile (\figref{fig:BoneProf}b) shows a tendency to overshoot for thicker regions and undershoot for thinner regions. On closer inspection of \figref{fig:BoneIm}b,e, it seems that the NearBRT reconstruction is sharper and more consistent, whereas VERT shows some variations in thickness and velocity. This is potentially correlated with the regions where the first arrival transitions between the through-transmission wave and the guided wave, so there may be a trade-off between obtaining useful information about the cancellous interior and cortical exterior. 

The cortical thickness of both VERT and NearBRT are thicker than the GroundTruth when evaluated from the half peak height \figref{fig:BoneProf}. As mentioned earlier, the first derivative regularisation added during the inversion step (see Section \ref{sec:inversion}) contributes to the blurring of both the periosteum and endosteum boundaries. The cortical thickness is always slightly thicker in VERT than NearBRT, which is consistent with the error introduced by the finite excitation spot discussed above for Smiley80. 

\section{Discussion} \label{sec:Discussion}
The results show that VERT has achieved the goals of higher extended-range resolution and high-contrast robustness. They also show that 1. the \textit{a priori} information supplied to VERT, 2. the virtualisation step, and 3. the initialisation step are all invaluable to solving high-contrast extended-range problems.

VERT's Bone200 reconstruction surpassed expectations in terms of precision and similarity despite the presence of guided waves within the cortical bone. Firstly, the eikonal equation does not model the interference or reflection of partial waves in waveguides. Secondly, the approximation of $c_{phase}=c_{group}$ in the eikonal equation \pareneqref{eq:eikonal} is not exact for the dispersive guided waves. The current work thus provides a basis for further analysis by first establishing the approximate thickness and material properties of the waveguide, which can be utilised as a starting point for guided-wave-specific algorithms. Our results have shown that it may be possible to utilise the virtual transducers (from the virtualisation step in VERT) as a phased array to excite specific guided wave modes from the ROI boundary, given the strength of such modes even when not intentionally exciting them.

Although arrival time selection was not the focus of the current study, there was some difficulty in its execution due to artefacts from 1. imperfect destructive interference and inexact focusing in the virtualisation step, as well as 2. the presence of guided waves. More sophisticated automated picking algorithms, such as those discussed in \cite{Pierre}, may be employed in future implementations. This is especially helpful if guided wave arrivals can be differentiated from through-transmission arrivals. The transmission of a chirp signal may help to this end. 

This work demonstrates that, while arrival time picking is a critical element of successful BRT imaging (as suggested by \cite{Pierre}), the exact form of the BRT problem is of vital importance. The approach of placing the array on the ROI boundary and managing the high contrast suitably enables BRT to achieve accurate results for these types of problems. The demonstration of VERT has also shown that this transducer placement can be achieved virtually, making it a practical solution when access is limited.

In the current work, validation was carried out with an \textit{a priori}, homogeneous ER. The resolution and robustness gains of VERT in this context show a clear demand for VERT to be extended to an \textit{a priori}, inhomogeneous ER. In addition, we have established that the \textit{a priori} information, including the ER velocity and the location of the ROI boundary, are critical to the success of any algorithm. 

In the context of \textit{in-vivo} femoral neck imaging, there could be other inhomogeneities in the ER, including various bone protrusions (e.g. greater trochanter, pelvic bone) just outside the imaging plane. VERT is much less susceptible to these high-contrast scatterers because of the 3D focusing in the virtualisation step, which results in little energy entering the protrusions, increasing the signal-to-noise ratio. Any affected transducer pairs can also be excluded from the physical FMC data before virtualisation. Scatterers close to the femoral neck could also be incorporated within the ROI boundary for detailed, assumption-free imaging. 

In real bone, the interface at the endosteum (between the cortical and cancellous bone) is not a clean smooth boundary as in Bone200 (\figref{fig:SampleProblems}b). The true endosteum has protruding trabeculae (which form part of the cancellous matrix), acting as scatterers that deflect energy from the guided wave. Given that the eikonal forward model in the BRT inversion step does not model guided wave effects, the results of VERT with real cancellous bone may prove to be even better. In addition, the real cancellous bone would not be homogenised as it is in the current paper, such that the Biot fast wave that enters from the cortical bone radially into the cancellous bone may experience a lower acoustic contrast, further strengthening the through-transmission.

In a medical setting, the conservatively chosen 200mm array for Bone200 will only serve the lowest 5th percentile of females across ethnicity and age groups \parencite{Anthropometric}. To cover the 95th percentile ER diameter, a much larger array diameter of 300mm will be needed. Given its high robustness and extended-range resolution, VERT will even more dramatically outperform BRT when translated to \textit{in-vivo} applications.

VERT also has the potential to replace classical BRT in many current BRT applications in most biomedical and industrial applications, such as breast imaging and guided wave corrosion patch tomography. 
The acquisition array in these applications is usually oversized to cover the largest ROIs. 
This results in a large effective ER-to-ROI size ratio for the majority of ROIs, despite the lack of an actual ER impeding access to the ROI boundary. This gives VERT a huge advantage over classical BRT.

\section{Conclusions}
BRT is one of the most robust and fastest tomographic algorithms, so it is widely used in tomography. However, even BRT struggles with high-contrast extended-range tomography, where transducers have no direct contact with the high-contrast region of interest (ROI). \textit{In-vivo} femoral neck imaging is one such problem, desperately needed by millions of osteoporotic patients to predict hip fracture risk. 

We have introduced VERT which increases BRT's high-contrast robustness and extended-range resolution. VERT simplifies the problem by utilising \textit{a priori} information to segment the problem into two major regions: the unknown ROI and the known external region (ER). Two additional steps are introduced: virtualisation and bi-velocity initialisation. The virtualisation step involves placing virtual transducers on the minimum convex boundary around the ROI, from which BRT-style inversion can be employed. We showed mathematically that this unlocks high resolution that is unachievable with BRT from afar. The virtualisation step also increases robustness through increasing ray sampling density within the ROI and unlocks the initialisation step which increases robustness by rapidly estimating the overall properties of the ROI before inversion.

VERT was validated through two 2D sample problems \textit{in silico}, Smiley80 and Bone200. Smiley80 is a smiley-inscribed cylinder with bone-like properties within soft tissue; Bone200 is based on the setup of an \textit{in-vivo} bone imaging problem, with a simplified ellipsoid bone model surrounded by soft tissue. The ROI includes a cortical bone exterior that varies in thickness and a cancellous bone interior that varies in material properties. 

VERT has drastically improved resolution and robustness when compared to both BRT and a best-case scenario of BRT (BestBRT) while maintaining or even improving speed. 
This improvement becomes pronounced as the ER becomes much larger than the ROI. 
Furthermore, we demonstrated that VERT is approaching the resolution achievable from direct access to the ROI boundary (NearBRT), despite having no contact with the ROI in practice. Some minute resolution gains may still be possible to this end, but it is found that some of the minor focusing errors in the virtualisation step proposed may benefit VERT in solving the inversion problem. 

VERT has shown that high-contrast extended-range tomography problems are solvable without any priors about the ROI interior. 
With further work, VERT shows promise in achieving \textit{in-vivo}, site-specific, quantitative bone health imaging, which will be useful in both diagnosis and monitoring of osteoporosis. 

VERT also has the potential to replace classical BRT in many other applications, given the drastic resolution gains where the physical array diameter is large. 
For further resolution gains, VERT provides an excellent and trustworthy starting background for algorithms with less robustness and high computational costs.

\printbibliography
\end{document}